\documentclass[conference]{IEEEtran}
\IEEEoverridecommandlockouts
\usepackage{amsmath, amsfonts}
\usepackage{algorithmic}
\usepackage{algorithm}
\usepackage{array}
\usepackage{float}
\usepackage{textcomp}
\usepackage{stfloats}
\usepackage{url}
\usepackage{verbatim}
\usepackage{graphicx}
\usepackage{cite}
\usepackage{pifont}
\usepackage{multirow, multicol}
\usepackage{adjustbox}
\usepackage{hyperref}
\hypersetup{
  colorlinks   = true, 
  urlcolor     = blue, 
  linkcolor    = blue, 
  citecolor   = green 
}

\begin{document}
\title{Design of JiuTian Intelligent Network Simulation Platform}

\author{
\IEEEauthorblockN{1\textsuperscript{st} Lei Zhao}
\IEEEauthorblockA{
\textit{China Mobile Research Institute}\\
Beijing, China \\
zhaoleiai@chinamobile.com}
\and
\IEEEauthorblockN{2\textsuperscript{nd} Miaomiao Zhang}
\IEEEauthorblockA{
\textit{China Mobile Research Institute}\\
Beijing, China \\
zhangmiaomiao@chinamobile.com}
\and
\IEEEauthorblockN{3\textsuperscript{rd} Guangyu Li}
\IEEEauthorblockA{
\textit{China Mobile Research Institute}\\
Beijing, China \\
liguangyu@chinamobile.com}
\and
\IEEEauthorblockN{4\textsuperscript{th} Zhuowen Guan}
\IEEEauthorblockA{
\textit{China Mobile Research Institute}\\
Beijing, China \\
guanzhuowen@chinamobile.com}
\and
\IEEEauthorblockN{5\textsuperscript{th} Sijia Liu}
\IEEEauthorblockA{
\textit{China Mobile Research Institute}\\
Beijing, China \\
liusijia@chinamobile.com}
\and
\IEEEauthorblockN{6\textsuperscript{th} Zhaobin Xiao}
\IEEEauthorblockA{
\textit{China Mobile Research Institute}\\
Beijing, China \\
xiaozhaobin@chinamobile.com}
\and
\IEEEauthorblockN{7\textsuperscript{th} Yuting Cao}
\IEEEauthorblockA{
\textit{China Mobile Research Institute}\\
Beijing, China \\
caoyuting@chinamobile.com}
\and
\IEEEauthorblockN{8\textsuperscript{th} Zhe Lv}
\IEEEauthorblockA{
\textit{China Mobile Research Institute}\\
Beijing, China \\
lvzheyjy@chinamobile.com}
\and
\IEEEauthorblockN{9\textsuperscript{th} Yanping Liang}
\IEEEauthorblockA{
\textit{China Mobile Research Institute}\\
Beijing, China \\
liangyanping@chinamobile.com}\
}

\maketitle

\begin{abstract}
This paper introduced the \textit{JiuTian Intelligent Network Simulation Platform\footnote{\href{jiutian.com}{JiuTian Intelligent Network Simulation Platform}}}, which can provide wireless communication simulation data services for the Open Innovation Platform. The
platform contains a series of scalable simulator functionalities, offering open services that enable users to use reinforcement learning algorithms for model training and inference based on
simulation environments and data. Additionally, it allows users to address optimization tasks in
different scenarios by uploading and updating parameter configurations. The platform and its open
services were primarily introduced from the perspectives of background, overall architecture,
simulator, business scenarios, and future directions. 
\end{abstract}


\section{Introduction}
\label{intro}
The application of artificial intelligence technology to solve technical and application problems in the communication field and vertical industries has become a mainstream direction\cite{you2019ai, schank1991s, li2017intelligent}. However, the communication network system is robust, complex, and has a long industrial chain. The intersection of communication and AI poses significant challenges. Additionally, the communication industry lacks a flexible and adaptable real-world verification environment, which hinders the iterative testing for the development and validation of network AI capabilities.

To address these challenges, the JiuTian Intelligent Network Simulation Platform(JINSP) provides an implementation solution for innovative smart networks, offering services in four dimensions: dynamic network simulation, model training, capability invocation, and capability evaluation. By creating virtual objects in the digital world to represent communication network entities, people, network services, and their topological relationships, it can effectively simulate, respond to, and predict the states and behaviors of various entities in the physical environment. This facilitates the incubation and breakthrough of network intelligence technology.

The JINSP provides various types of simulation services for research and application personnel in the "AI + communication" field and releases corresponding research tasks. The platform aims to simulate and construct a simulation environment for wireless communication networks, industry-specific networks, etc., of a certain scale. This environment consists of different business scenarios and can serve scenarios such as intelligent network operation, maintenance, optimization, and service provision, as well as the development, testing, and validation of AI capabilities for network element intelligence.

\section{Related work}
\label{related}
The JINSP follows the design principles and implementation logic of some commonly used AI platforms and simulation software in the industry. It is dedicated to providing an open-service platform for the wireless field. The following is a comparative analysis between the intelligent network simulation platform and several types of platforms currently existing in the industry.
\begin{itemize}
    \item \textbf{AI General Cloud Services:} This type of platform's service capabilities are mostly built on its own core business and provide a wide range of interface-based applications for the general audience. Users can obtain inference results by calling the API of the AI cloud service platform. Some well-known examples include Google Cloud AI Platform, OpenAI API, Baidu AI Open Platform, Tencent Cloud AI Lab, and so on. The original intention of the intelligent network simulation platform design is also based on existing core businesses, aiming to achieve platformization of services, and support user interaction and capability openness.
    \item \textbf{Online Training Platforms:} This type of platform supports mainstream algorithm frameworks and provides GPU computing resources as well as some task datasets. Its capabilities cover academic areas such as deep learning, computer vision, natural language processing, etc.
    
    This category of platforms is favored by educators and researchers due to its superior hardware environment and user-friendly interface. However, these platforms primarily serve to provide GPU computing services, and only a small portion of datasets applicable to communication directions. They have a limited impact on research in the "AI + communication" field and still cannot address the issues of limited data and lack of interactive network environments in communication research.
    \item \textbf{Toolkits:} This type of platform does not rely on interfaces for online inference, nor does it provide a trainable platform. Instead, it offers the complete core product and functionality to the user. Users can download or reference the toolkit to create new instances locally, with each instance tailored to meet individual user needs. Users can then use this as a basis for training or inference. For example, with OpenAI's Gym\cite{goulao2023gymnasium} platform, developers can easily build and train intelligent agents \cite{bellemare2013arcade, kaelbling1996reinforcement, schulman2017proximal} and evaluate their performance.
    \item \textbf{Simulation Platforms:} This category of platforms focuses on the simulation implementation in specific vertical domains and does not inherently include AI capabilities.\cite{guo2020gluoncv} The simulation effects are achieved through underlying engines. Some examples include Atoll, Planet, Exata, MATLAB, etc. Atoll and Planet specialize in specific types of network simulation, such as wireless network planning and satellite communication system design. Exata supports various network types and technologies, including wireless networks and sensor networks. These platforms are suitable for professional users in related fields. However, due to the lack of support for AI algorithms, their scalability and user flexibility are relatively low.

    MATLAB provides a rich set of mathematical and simulation toolboxes for various simulation modeling and performance evaluation tasks. It also offers the capability to incorporate extended algorithm functionalities. However, using MATLAB as a platform can be more challenging as it requires a corresponding programming background.
\end{itemize}

Compared to the various open platforms described above, the intelligent network simulation platform designed in this paper integrates features such as online training and inference capabilities, dynamic simulation functionality, and future open API services. Additionally, it is capable of providing specific AI capabilities for the wireless communication vertical domain's particular business needs. 

\section{Framework}
\label{framework}
\begin{figure*}
    \centering
    \includegraphics[width=0.95\linewidth]{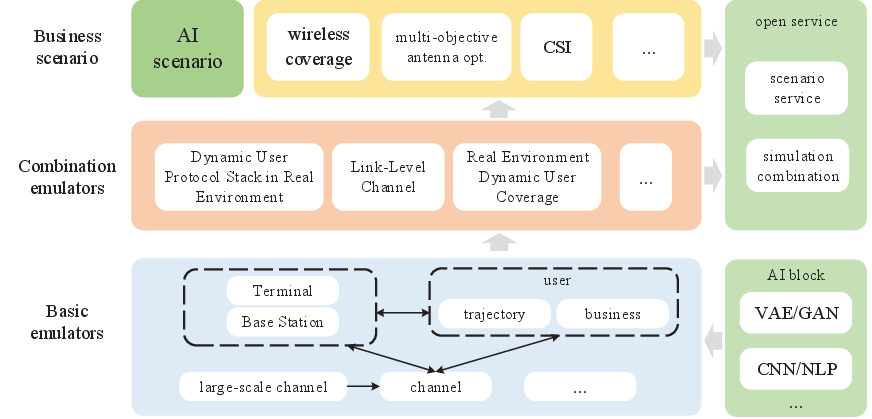}
    \caption{The Overall Architecture Diagram of the Intelligent Network Simulation Platform, includes basic emulators, combination emulators and business scenario. Besides, open service and AI block can provide the basic service for the JINSP.}
    \label{fig:framework}
\end{figure*}

The JINSP constructs different types of simulators and integrates them based on the requirements of business scenarios. This enables users to conduct online model training and inference for corresponding open tasks. Users immerse themselves in the simulation platform environment and interact with it to invoke simulation capabilities and manage tasks. For instance, users can issue simulation configurations to invoke specific simulation capabilities and perform simulation functions. Subsequently, users can update simulation configurations based on the simulation results returned by the environment (e.g., antenna parameters, base station positions, etc.). They can iterate the invocation of simulator capabilities to accomplish online model training and inference. This process is similar to the concept of an intelligent agent \cite{henderson2018deep, iwamura2010carrier}, which can adapt its understanding based on the environmental configuration. This chapter will provide a detailed introduction to the basic simulators and the process of combining them in Figure \ref{fig:framework}.

\subsection{Basic Emulator}
\begin{itemize}
    \item \textbf{User Behavior Simulation} involves trajectory generation and business generation. Trajectory generation utilizes generative algorithms such as Variational Autoencoders (VAE) \cite{long2023practical, feng2020learning, kingma2013auto} and Generative Adversarial Networks (GAN) \cite{radford2015unsupervised, goodfellow2020generative} to simulate user trajectories, achieving the generation of latitude and longitude sequences at a daily level. Business generation addresses the synthesis of user-level business traffic. It is based on a Knowledge-Enhanced GAN model. This model takes user behavior representation vectors and packet distribution characteristic vectors extracted from user traffic data packets as inputs for knowledge enhancement. Subsequently, it employs GAN to generate packet sequences.
    \item \textbf{Large-Scale Channel Simulation} is based on the antenna pattern of any sub-beam, combined with the location, antenna height, transmit power, mechanical azimuth, tilt angle, geographical terrain, etc., simulate the large-scale channel simulation results for each grid or user-specific cell with any type of beam. Subsequently, this supports the calculation of relevant metrics for base stations and terminals, such as Service Cell RSRP/SINR \cite{afroz2015sinr, park2016analysis} values, neighboring cell RSRP/SINR values, and the calculation of the corresponding beam IDs for the service cell and neighboring cells.
    \item \textbf{Channel Simulation} is based on real geographical environments and considering the regional characteristics of the simulation scenario (such as Indoor, Umi, Uma, etc.), wireless channel modeling is conducted across regions and multiple scenarios. Channel simulation achieves channel modeling by invoking the foundational simulator, which includes Large-Scale Channel Simulation, and combines it with small-scale fading (fast fading) simulation. In this process, small-scale fading simulation employs a common statistical modeling approach, thoroughly simulating the multipath effects generated by processes such as direct propagation, reflection, diffraction, transmission, and diffuse scattering in the transmission of electromagnetic waves. Finally, the results of the large-scale fading model are combined with those of the small-scale fading model to generate the simulated channel matrix.
    \item \textbf{Base Station/Terminal Simulation} provides wireless protocol stack processes and functional simulation for Radio Resource Control (RRC), Medium Access Control (MAC), and Physical Layer (PHY) in system-level simulation \cite{mountaser2017feasibility}.

    In the RRC layer simulation, key processes like cell access, handover/reselection, and wireless resource management are included. The MAC layer simulation, conducted separately on both the base station and terminal sides, models critical processes such as wireless resource scheduling, MIMO \cite{larsson2014massive}, link adaptation, wireless resource mapping, uplink power control, and Hybrid Automatic Repeat reQuest (HARQ). The PHY layer simulation, guided by the functional processes in the above two layers, ultimately calculates system-level network performance metrics like the number of connected users, cell load, uplink/downlink transmission rates, and more.

    Additionally, the Base Station/Terminal Simulation module can be combined with map information, cell parameters, and beam configurations. With the aid of AI-calibrated path loss calculation formulas, it computes the Service Cell RSRP/SINR values under any sub-beam antenna pattern in wireless coverage simulation scenarios.
\end{itemize}

\subsection{Combination Emulator}
In the previous section, individual basic simulators are responsible for simulating the fundamental capabilities within the network system, but they may not meet the simulation requirements of complex business scenarios \cite{maeder2014towards}. The platform combines multiple basic simulators and provides unified services to meet the task opening and online model training requirements of different business scenarios. Currently, the intelligent network simulation platform provides three types of combination simulators:
\begin{itemize}
    \item Real Environment Dynamic User Protocol Stack Simulation: This combination simulator is composed of user behavior simulation, base station simulation, and terminal simulation. It provides users with metrics such as RSRP, SINR, traffic, and rates at both user and cell granularity in a real environment.
    \item Real Environment Dynamic User Coverage Simulation: To improve the efficiency of base station and terminal simulation, this simulator primarily invokes the physical layer simulation from the above simulators. It combines with user simulation to output coverage metrics such as RSRP and SINR at both user and cell levels.
    \item Link-Level Channel Simulation: This combination simulator utilizes the core capability provided by channel simulation. It combines with base station and terminal physical layer protocol stack simulation to provide users with frequency-domain channel response information at the Resource Element (RE) level.
\end{itemize}

\section{Business Scenario}
\subsection{Multi-objective antenna opt.}
Multi-objective optimization aims to assist network operators in using the simulation platform's twin-modeling capability of real network situations. This allows for a cost-effective understanding of the evaluation results of various network metrics under different cell SSB antenna weight parameters \cite{guo2020weighted}. Consequently, it enables more convenient and efficient optimization results in different regions and time periods, considering multiple optimization objectives.

A wireless cellular network consists of multiple base stations, each with several cells. These cells achieve coverage through the transmission of Synchronization Signal Blocks (SSBs). In 5G, cells can adjust the azimuth, tilt angle, horizontal beamwidth, and vertical beamwidth of eight sub-beams to control the coverage range and radius of the cell. Users scan through each SSB beam and select the one with the strongest coverage signal. The cell simulates access metrics and rates based on factors such as the user's coverage signal and interference.

For the open task of multi-objective antenna optimization, considering user mobility, the task involves adjusting the sub-beams of the cell, including azimuth, tilt angle, horizontal beamwidth, and vertical beamwidth. The objective is to jointly optimize the metrics \cite{kingma2014adam, poli1993use} of all cells within the region, allowing for the optimization of SS-RSRP, SS-SINR, total number of users, total traffic volume, and rates within a five-minute granularity. This optimization task aims to achieve the overall best performance in the region, encompassing $n$ cells and $k$ users. Additionally, by controlling the weighting coefficients, the optimization task can be tailored to focus on specific metrics, transforming the problem into a coverage optimization problem. The output of this task includes the azimuth, tilt angle, horizontal beamwidth, and vertical beamwidth of each cell within the region every five minutes, ensuring that various metrics (as mentioned above) reach their optimal values across the entire area within a specified time frame.

\subsection{CSI}
In large-scale MIMO systems, base stations (BS) are typically equipped with as many as several hundred active antennas, serving multiple user equipments (UEs). Accurate Channel State Information (CSI) \cite{guo2022overview} is crucial for harnessing the performance gains of large-scale MIMO. The CSI for the downlink channel needs to be obtained by UEs through downlink pilot estimation and then transmitted back to the base station through a feedback link. The base station utilizes the received feedback information for tasks like precoding and other adaptive transmission optimizations. Deep learning-based methods can address the CSI compression feedback issue in large-scale MIMO systems, efficiently and accurately enhancing the feedback precision. There are also many promising research directions in this area worthy of further investigation.

This task aims to harness the feature extraction and information compression capabilities of AI. It involves training an AI model using channel characteristic information. This model is then used to compress the channel characteristic information at the UE side. The compressed information is transmitted through the channel and recovered at the receiving end. The base station uses the recovered information for tasks like precoding and other adaptive transmission optimizations.

The ultimate goal of CSI compression feedback is to restore the channel state information with as little loss as possible under a fixed compression bit requirement. Smaller compression bits result in fewer resources being used for transmission. However, this can lead to lower precision in model recovery, resulting in lower feedback performance. This task requires designing a model under a specific feedback bit vector, with the objective of achieving higher restoration accuracy of the model with the smallest possible compression bits.

\section{Future decision}
\label{future}
The platform starts from the interdisciplinary integration of communication and artificial intelligence, selects key technologies and application issues of concern to the academic and production fields, analyzes and extracts the necessary simulation environment capabilities for algorithm technology development corresponding to various problems. It has constructed an intelligent network simulation service that integrates various functions such as user behavior simulation, wireless network coverage simulation, base station protocol stack simulation, and channel simulation, forming a systematic network simulation environment. Through simulation performance optimization, capability combination, and interface encapsulation, the platform provides a good user experience for the external service opening of simulated network environments and data. It also contributes important infrastructure to the integration and development of communication and artificial intelligence technologies.

To provide stronger support for technological breakthroughs and promote industry development, the platform still needs to continuously accumulate strength in the following directions: First, with the development of generative artificial intelligence and large-scale model technologies, artificial intelligence technology has gradually made significant breakthroughs in many areas that were once considered unsuitable or unfeasible. The future focus of the platform is how to use multimodal communication network big data and large-scale model technology to build network simulation capabilities. Second, digital twinning is an important evolution direction of 6G communication technology. The platform will continue to expand the business scenarios covered by simulation capabilities, achieve end-to-end full-link simulation, and accelerate the construction and opening of trial networks in existing networks. It will overcome the coexistence of simulation environment and real network, providing a reliable basis for the technical feasibility demonstration of digital twin networks. Third, to meet the personalized research and production needs of users from all walks of life, the platform will continuously improve the flexibility and freedom of the network simulation environment. It will support more parameters configuration such as user behavior, network equipment, and simulation models, while actively promoting the decoupling of simulation capabilities and the design of unified interfaces. It will mobilize industry partners to contribute basic simulation capabilities and build a flexible and assembled network simulation environment. It will promote the open source and open access of high-quality resources in the industry, establish a collaborative and innovative industry ecosystem, and provide strong support for the national network power strategy.

\bibliographystyle{IEEEtran}
\bibliography{ref.bib}

\end{document}